\documentclass[]{llncs}

\usepackage{wrapfig}
\usepackage{graphicx}
\usepackage{amsmath}

\newtheorem{Ex}{Example}
\newtheorem{tr-rule}{Rule}

\title{L2C2: Logic-based LSC Consistency Checking}
\author{Hai-Feng Guo, Wen Zheng, Mahadevan Subramaniam}
\institute{
        Department of Computer Science\\
        University of Nebraska at Omaha\\
        Omaha, NE 68182-0500, USA\\
        Emails: \{haifengguo,wzheng, msubramaniam\}@mail.unomaha.edu
}

\date{}			
\begin{document}
\thispagestyle{empty}
\maketitle


\begin{abstract}
Live sequence charts (LSCs) have been proposed as
an inter-object scenario-based specification and
visual programming language for reactive systems.
In this paper, we introduce a logic-based framework to
check the consistency of an LSC specification.
An LSC simulator has been implemented in logic programming, utilizing a memoized
depth-first search strategy,
to show how a reactive system in LSCs would response to
a set of external event sequences.
A formal notation is defined to specify external event
sequences, extending the regular expression with
a parallel operator and a testing control.
The parallel operator allows interleaved parallel external events
to be tested in LSCs simultaneously;
while the testing control provides users
to a new approach to specify and test certain temporal properties
(e.g., CTL formula) in a form of LSC.
Our framework further provides either a state transition graph or
a failure trace to justify the consistency checking results.
\end{abstract}

\section{Introduction}

Live sequence charts (LSCs)~\cite{DH99,HM03} have been introduced
as an inter-object scenario-based specification and visual programming
language for reactive systems. The language extends  traditional sequence charts,
typically sequence charts in UML~\cite{UMLdocs} and
message sequence charts (MSCs)~\cite{Z120}, by
enabling mandatory behaviors and forbidden behaviors specification.
With mandatory behaviors, LSCs are able to specify necessary
system behaviors in a more semantically precise manner instead
of only posing weak partial order restriction on possible behaviors
as in sequence charts. Such a semantical precision is critical to
upgrading the LSCs from a formal specification tool to
an achievable dream of becoming a scenario-based programming language~\cite{H00},
since mandatory behaviors exactly tell what
to expect at system runtime. In addition, a forbidden behavior
specification further enriches the LSCs with self-contained consistency checking
capability; it allows the LSCs to specify the failure scenarios
at runtime by checking reachability instead of
testing failure conditions explicitly and blindly at every
running step. LSCs have been successfully used in
many real-life applications such as hardware
and software verification~\cite{BGS05,CHK05},
an air traffic control system~\cite{BHK03}, and a radio-based train system~\cite{BDK02}.
The features of specifying mandatory and
forbidden behaviors have recently been incorporated into
the UML 2.0~\cite{UML2} to increase its expressive power.

Consistency checking~\cite{HK99,KH05,SD05,KTWW06,KM07,KM08}
is one of the major and formidable problems on LSCs.
In a complicated system consisting of many LSCs,
inconsistency may be raised by inherent contradiction
among multiple charts or due to inappropriate environmental/external
event sequences. Previous work on consistency checking
are mainly focused on automata-based strategies.
In~\cite{HK99}, the consistency of LSCs is shown as
a necessary and sufficient condition for the existence
of an object system satisfying it; the consistency checking
is then reduced to a problem whether a satisfying object
system, in a form of automaton, can be synthesized from LSCs.
Other automata-based transformation includes~\cite{BH01,KTWW06,KM08},
which turn the LSC specification into variants of
B\"{u}chi automata for verification.
In~\cite{SD05}, a semantic transformation from
LSCs to communicating sequential process (CSP) is presented,
so that the consistency checking of LSCs can be done by reusing
an existing tool for CSP. There has been some other work done
in translating LSCs to temporal logic for
model checking~\cite{KH05,KM07}. All these automata-based
consistency checking strategies lack the capabilities
for LSCs users, at the language level, to specify any
user-preferred testing and simulate running LSCs in any way.
Therefore, they can only be served as a
formal verification tool to support static analysis on LSCs,
and at the same time, may suffer the complexities
caused by transformation itself, automata synthesis, or
the blowing size of transformed results~\cite{TW06,HMS08}.

In this paper, we introduce a logic-based framework to
implement an automated LSC simulator with practical user controls,
and to check the consistency of an LSC specification with
sufficient justification. An LSC simulator has been implemented
in logic programming to show how a reactive system in LSCs would
response to a set of external event sequences.
A formal language is defined for users to specify external event
sequences by extending a regular expression notation with
a parallel operator and a testing control. The parallel operator
allows the LSC simulator to test the scenarios where multiple external
events may happen simultaneously; and the testing control provides
users with a new approach to specify certain temporal properties
(e.g., CTL formula~\cite{CGP01}) in a form of LSC.

We present a new high-level computational (operational)
semantics of LSCs to show how a running LSC affects the system
behaviors in response to a continuous input of external events.
The semantics is defined in the form of a derivation tree,
called {\em PLAY-tree}, where each branch from the root to
a leaf corresponds to a possible LSC run on a finite sequence of
external event inputs. The consistency of an LSC specification
$\mathcal{L}$ is defined as, given a nonempty language
$\mathcal{I}$ of external events, whether there exists a
corresponding {\em PLAY-tree} with all successful branches.
If such a PLAY-tree exists, then the LSC $\mathcal{L}$
is consistent on the input language $\mathcal{I}$;
otherwise, a failure trace can be obtained along
the failure branch in the PLAY-tree.
As a result, the consistency checking of an LSC is basically
an ordered traversal of the PLAY-tree.
Our LSC simulator utilizes a traversal algorithm using
a memoized depth-first search strategy for
efficient consistency checking of LSCs.

Besides running LSCs and checking their consistency, our
LSC simulator further generates evidences to justify the
truth value of checking results. If the consistency is
true, the simulator can generate a state transition
diagram to illustrate system behaviors as well as
CTL formula satisfaction. Otherwise, if the checking
result is false, the LSC simulator returns a least prefix
failure trace as a counter-example evidence, so that the
LSC users can easily re-construct the simulation anti-scenarios.

In this paper, we adopt assumptions similar to~\cite{HM03}.
We assume all messages in LSCs are synchronous
and that no sending messages are lost over communication channels;
the running LSCs use an input-enabled concurrency model, where
no two external events occur at the same time and
that internal events are processed at real-time,
much faster than receiving a next external event.
It needs to be mentioned that even though we
assume no simultaneous external events in our LSC simulation,
interleaved parallel external events can still be simulated
with extra language features as described in later sections.

The paper is structured as follows. Section~\ref{sec:lsc}
gives a brief introduction of the syntax and semantics of
LSCs through a web order example.
Section~\ref{sec:l2c2} presents an architecture of
a logic-based LSC consistency checker.
Section~\ref{sec:l2s} illustrates how a running LSC reacts to the
environment events using a computational tree, and shows
how the consistency of LSCs can be achieved through a memoized depth-first
search strategy.
Section~\ref{sec:eesl} defines a formal language
for specifying the external events with parallel and testing controls.
Section~\ref{sec:justification} shows the supporting evidence for
LSC consistency checking and simulation.
Finally, conclusions are given in Section~\ref{sec:conclusion}.

\section{Live Sequence Charts (LSCs)}
\label{sec:lsc}

LSCs have two types of charts: {\em universal}
and {\em existential} charts. A universal chart is used to
specify a scenario-based {\em if-then} rule, which
applies to all possible system runs. A universal chart
typically contains a {\em prechart}, denoted by a top dashed hexagon,
and a {\em main chart}, denoted by a solid rectangle right below
a prechart;
if the prechart is satisfied, then the system is forced to
satisfy the defined scenario in the main chart.
An existential chart is usually used to specify a testing scenario
that can be satisfied by at least one possible system run.
In this paper, we will focus on universal charts only, since
achieving the consistency among universal charts on all system runs
is more interesting and difficult; in addition,
a testing scenario can easily be specified in a universal chart in
our framework.

\begin{figure*}[!ht]
	\centering
	\begin{tabular}{cc}
		\hspace{-.2in}
		\includegraphics[width=2.4in,height=1.6in]{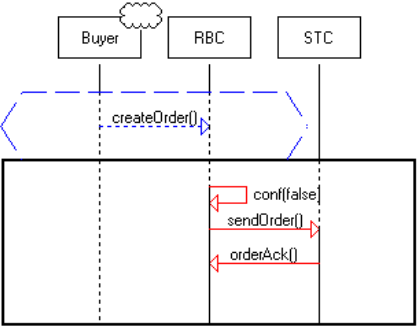} &
		 \includegraphics[width=2.4in,height=1.6in]{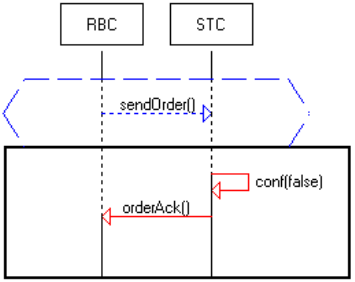}\\
		(a) RBC creates an order & (b) STC receives an order \\
    \includegraphics[width=2.4in,height=1.8in]{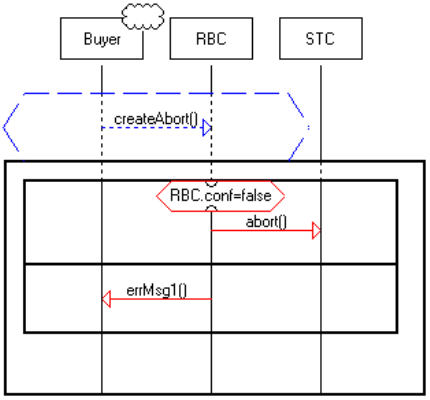}&
    \includegraphics[width=2.4in,height=1.8in]{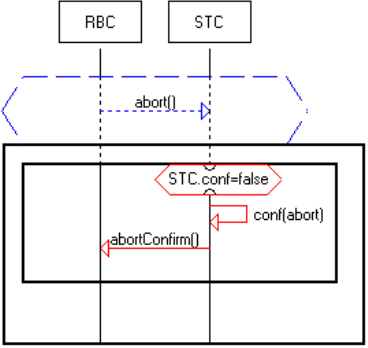}\\
		(c) RBC aborts the order & (d) STC receives an ``abort'' \\
		\raisebox{1.0in}[0pt]{
	  \begin{tabular}{c}
	  \includegraphics[width=2in,height=1.2in]{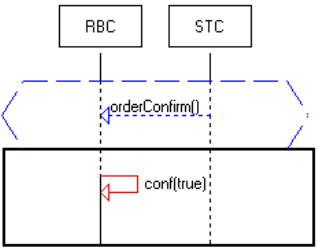} \\
	  (e) RBC receive an order confirmation
	  \vspace{0.1in}\\
	
	  \includegraphics[width=2in,height=1.2in]{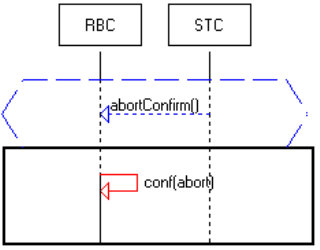}
	  \end{tabular}}
	   &
		\includegraphics[width=2.4in]{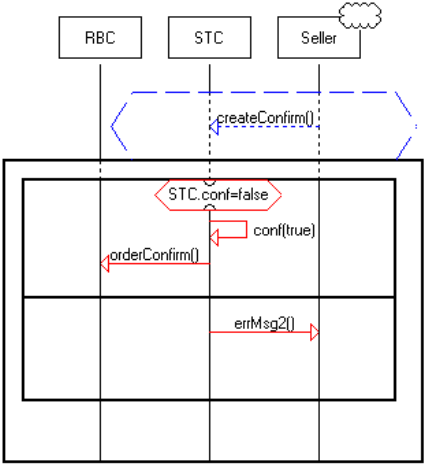} \\
		(f) RBC receive an abort confirmation & (g) STC confirms the order
	\end{tabular}
\caption{An LSC specification for a web order system}
\label{fig:webService}
\end{figure*}

Another main feature of LSCs is the assignment of temperatures,
{\em hot} or {\em cold}, to various elements in a chart including
locations, messages, and conditions. Within a chart, a hot element
signifies that things that must move on, while a cold element
signifies things that may occur. The combination of universal charts
and hot elements can be used to specify forbidden scenarios.

\subsubsection{A Web Order Protocol Example}

Figure~\ref{fig:webService} shows six universal LSCs
for modeling a web order protocol between
a really big corporation (RBC) and a small T-shirt company
(STC)\footnote{This web order protocol was originally posted at
petri-pi mailing list by N. Kavantzas on August 2005.}.
RBC and STC each has a state variable, {\em conf}, with the domain
{\tt \{false, true, abort\}} denoting whether the order has been initiated,
confirmed or aborted respectively.
LSCs (a)(c)(e)(f) illustrate four scenarios for the RBC.
Chart (a) shows if a buyer creates an order, then
the order is initiated in RBC by setting {\em RBC.conf} false,
sending a {\em sendOrder} message to STC, and waiting
for the acknowledgment; chart (c)
shows if the buyer wants to abort the order, and
at that point if the order has not been confirmed yet,
an abort message will be forwarded to STC, or otherwise,
an error message will be issued; and charts (e) and (f), respectively,
show that RBC receives an order or an abort confirmation from STC,
and then sets its state variable, {\em RBC.conf}, {\tt true} or {\tt abort}.
LSCs (b)(d)(g) describe behaviors from the STC point of scenarios.
Chart (b) says that if STC receives an order request,
it will set {\em STC.conf} {\tt false},
and then reply to RBC an acknowledgment;
chart (d) says that if STC receives an abort request,
and at that point, if {\em STC.conf} is still false,
then set it {\em abort} and reply with an abort confirmation message;
chart (g) says that if the seller wants to confirm the order,
and at that point if the order is not aborted or confirmed yet,
it will set {\em STC.conf} {\tt true} and send
a confirmation message to RBC.

\subsubsection{Semantics}

Now we briefly introduce the semantics of LSCs,
whose detailed explanation can be found in~\cite{HM03}.
Given an LSC $L$, a running copy of $L$ is defined as
$\langle L_r, Mode, Cut\rangle$, where $L_r$ is a copy of the
original chart,
$Cut$ is a legal tuple of $L$ representing
the runtime locations of the instances of $L_r$,
and $Mode$ is either {\em PreActive} or
{\em Active} denoting whether the current $Cut$
is in prechart or main chart respectively.
Let $E$ be a set of all system events including
external events, internal messages among system objects, or
hidden events defined in LSCs.
Thus, the operational semantics of an LSC specification
$Ls$ with respect to a system model $Sys$ can be defined as a state
transition system $Sem(Ls, Sys) = \langle \mathcal{S}, s_0, \Delta\rangle$.
$\mathcal{S}$ is the set of possible states,
A state $s \in \mathcal{S}$ is defined as
$\langle Q, \mathcal{RL}, B\rangle$, where
$Q$ is a set of system object states in $Sys$,
$\mathcal{RL}$ is the set of currently running copies of LSCs, and $B$ indicates
by {\em True} or {\em False} whether the state is a violating one;
$s_0 = \langle Q_0, \emptyset, False\rangle$ is the initial state, where
$Q_0$ is a set of initial system object states; and
the function $\Delta: \mathcal{S} \times E \rightarrow \mathcal{S}$ is
the set of allowed transitions.
Given a current state $s \in \mathcal{S}$ and a system event
$e \in E$, the transition $\Delta(s,e)$ returns an updated state after
processing one or more of the following actions:
activating a universal chart copy to $\mathcal{RL}$ if applies,
removing a running LSC from $\mathcal{RL}$ if complete,
advancing the cuts in each running LSC in $\mathcal{RL}$ by processing $e$
and updating object states in $Q$ accordingly if applies, or setting the flag
$B$ to {\em True} if violation.
Note that the definitions of a state transition system $Sem$ and its transition
function $\Delta$ are slightly different from the ones defined in~\cite{HM03}.
Here we include a system model $Sys$ and consider the changes of system object states
during the transition.

The Play-Engine in~\cite{HM03} provides an execution
platform for a given system model $Sys$ and its LSC specification $Ls$
by implementing the operational semantics $Sem(Ls, Sys)$.
The LSCs can be executed in phases of {\em step} and
{\em super-step}. The procedure for a step phase, given a transition system
state $s$ and a system event $e$, is essentially the application of $\Delta(s,e)$.
The super-step phase, given an external message $e$ at a super-step state,
continuously executes the steps associated with enabled internal events
until the system reaches a {\em stable} state where
no further internal or hidden events can be carried out.

A super-step state is either the initial state $s_0$ or a stable state after executing
an external event. Let $\mathcal{S}_s$ denote a set of all super-step states.
We introduce a new notation
$\nabla: \mathcal{S}_s \times \Sigma \rightarrow \mathcal{S}_s^{+}$
to denote the state transition related to a super-step procedure in~\cite{HM03},
which takes a super-step state $(Q,\mathcal{RL},B) \in \mathcal{S}_s$ and
an external message $a \in \Sigma$ as inputs and
returns a set of all possible next super-step states.
Multiple next super-step states are possible because the enable events
may contain interleaving events, which can be executed in a nondeterministic
order.

\section{The Architecture of L2C2}
\label{sec:l2c2}

\begin{figure}
	\centering
\includegraphics[width=5in]{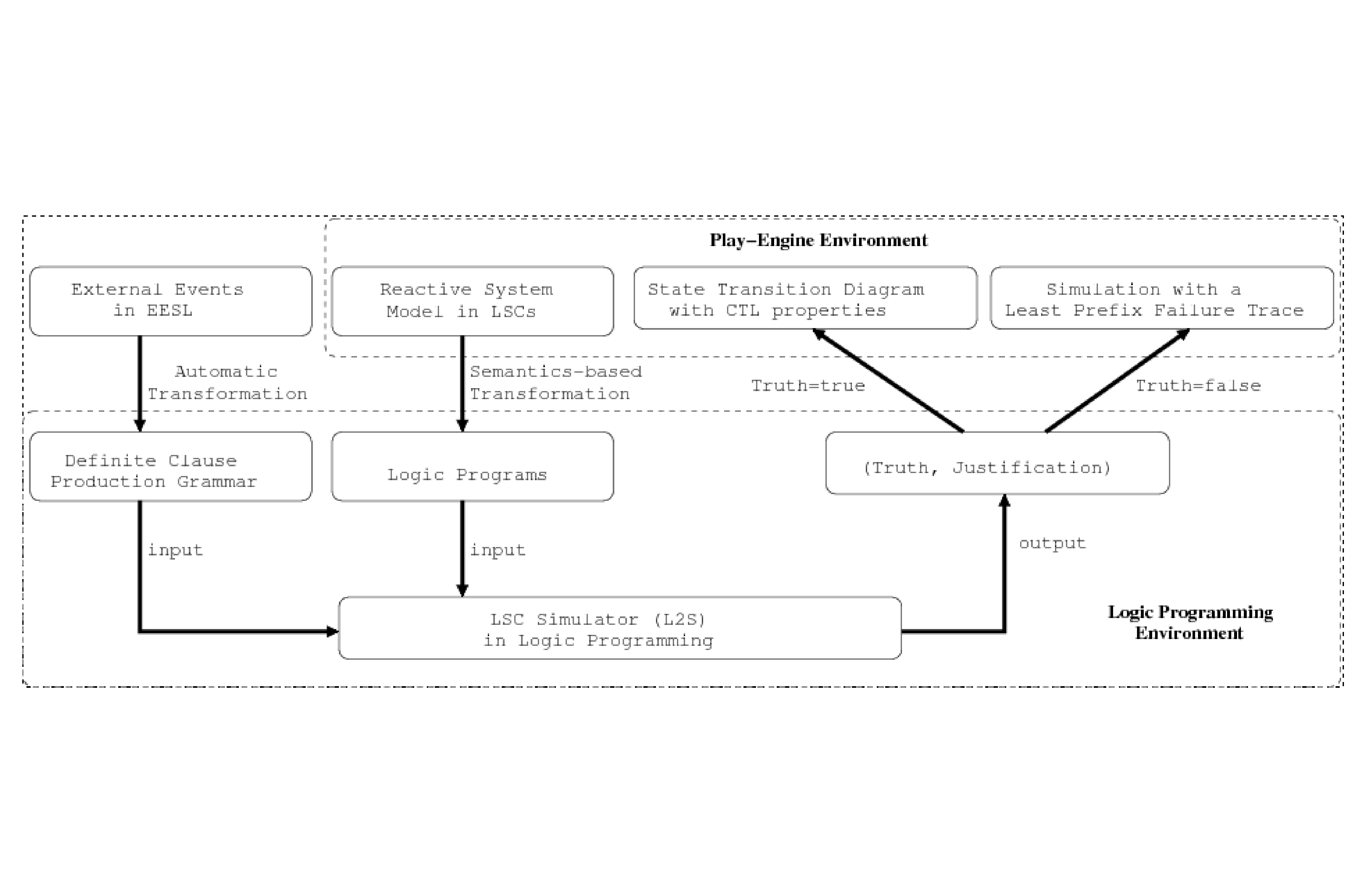}\\
\caption{An architecture of Logic-based LSC Consistency Checker (L2C2)}
\label{fig:l2c2}
\end{figure}

We have implemented a logic-based LSC consistency checking
system, named {\em L2C2}, using SWI-Prolog.
The architecture of our L2C2 tool, described in Figure~\ref{fig:l2c2},
shows an overview of the interaction between the scenario-based programming
tool, Play-Engine~\cite{HM03}, and a logic-based LSC simulator,
{\em L2S}. The L2C2 system, given an LSC-based model
of a reactive system and a specification
of external event sequences, verifies the consistency
of the LCS specification by simulating the running LSCs in
logic programming environment. The simulator {\em L2S} returns a
truth value as well as its justification, which provides simulation evidences
so that PLAY-engine users can easily follow the evidences to re-establish
the simulation scenarios in an interactive way. We adopted a logic programming
system due to its declarativity, strong symbolic reasoning capability,
and automatic backtracking.

The transformation from LSC specification to Prolog clauses adopts a
semantics-based approach.
The LSCs constructed in the PLAY-engine
are saved as XML files, which are processed by an XML parser written
in definite clause grammars (DCGs)~\cite{Abramson84} in Prolog. A semantic mapping function
is then defined to transform the parsing tree recursively to
logic clauses and maintain their semantic equivalence.
Both the parser and the semantic mapping function are encoded in Horn logic,
which results in an executable transformer directly.

\section{L2S: The logic-based LSC Simulator (L2S)}
\label{sec:l2s}

We present a running LSC simulator, named {\em L2S}, which takes inputs
an LSC specification, {\em Ls}, for a reactive system model {\em Sys},
and a context-free grammar (CFG)
$G = (\mathcal{V}, \Sigma, V_0, P)$
denoting a language $L(G) \subseteq \Sigma^{*}$,
and checks whether the running $Ls$ will react
consistently to any external event sequence $w \in L(G)$.
The computational semantics of our simulator essentially
extends the operational semantics defined in~\cite{HM03} with
the consideration of system object states and continuous
environment reaction.

\subsection{PLAY-tree}

A new succinct notation is introduced for describing the successive
configurations of the LSC simulator given the input of a CFG denoting
a set of an external event sequences.
A four-tuple $(Q, W, \mathcal{RL}, B)$,
where $Q$ is a set of current object states defined in $Sys$,
$W$ is the unprocessed part of the CFG in a sentential
form, $\mathcal{RL}$ is a set of current running LSCs,
and a boolean variable $B$ indicates by {\em True} or
{\em False} whether the state is a
violating one, is called an {\em instantaneous description (ID)}
of the simulator. The ID completely determines all the possible
ways in which the LSC simulator can proceed.

\begin{definition}[Move]
A {\em move} from one ID to another, denoted by the symbol $\vdash$,
could be one of the following cases:
\begin{itemize}
\item if $a$ is an external/terminal event, that is, $a \in \Sigma$ and $W \in \{\Sigma \cup \mathcal{V}\}^{*}$,
  \[
    (Q_1, aW, \mathcal{RL}_1, B_1) \vdash (Q_2, W, \mathcal{RL}_2, B_2)
  \]
  is possible if $(Q_2, \mathcal{RL}_2, B_2) \in \nabla((Q_1, \mathcal{RL}_1, B_1), a)$.
  Such a move is called a {\em terminal move}.
\item if $A$ is a nonterminal variable in $V$,
  \[
    (Q, AW, \mathcal{RL}, B) \vdash (Q, D_1\cdots D_nW, \mathcal{RL}, B)
  \]
  is possible if $A \rightarrow D_1\cdots D_n$ is
  a CFG production in $P$.
  Such a move is called a {\em nonterminal move}.
\end{itemize}
\end{definition}

Different terminal moves from a same ID are possible due to
inherent nondeterminism which could be caused by interleaving
messages in an LSC;
and different nonterminal moves from a same ID are also possible
due to multiple CFG productions from
the same variable to represent different message sequences. Therefore,
to show all possible moves systematically, we introduce
a derivation tree, called a {\em PLAY-tree}, to
illustrate the running LSC engine.

In a PLAY-tree, each parent-child edge represents an ID move. For clarity, we
add labels for each edge. For an edge of a terminal move, the label
is in form of {\tt [$a$]}, where $a$ is an external event triggering
the move; for an edge of a nonterminal move, the label is a production
rule $A \rightarrow D_1\cdots D_n$, which triggers the move.
Each branch of the PLAY-tree is a path from the root along a sequence of
edges. A {\em success branch} corresponds to a branch ending at a success leaf, $(Q, \lambda, R, False)$;
a {\em failure branch} corresponds to a branch ending at a violating leaf, $(Q, W, R, True)$;
and a branch with infinite moves is called an {\em infinite branch}.
For a finite branch, if we concatenate all the labels (external events)
of the terminal moves along the path from root to leaf, we call it
the {\em event sequence} of the branch.

\begin{Ex}
  Consider the web order example (Fig.~\ref{fig:webService}) with the
  external event input $(createOrder\cdot (createAbort + createConfirm))^{*}$.
  The regular expression can be represented in a context-free grammar $G$ with the main variable $W$ as follows:
  \begin{eqnarray*}
    W & \rightarrow & \lambda\ \mid\ createOrder\cdot AW \\
    A & \rightarrow & createAbort\ |\ createConfirm
  \end{eqnarray*}
\label{ex:webService1}
\end{Ex}

\begin{figure*}[htb]
	\centering
\includegraphics[width=5in]{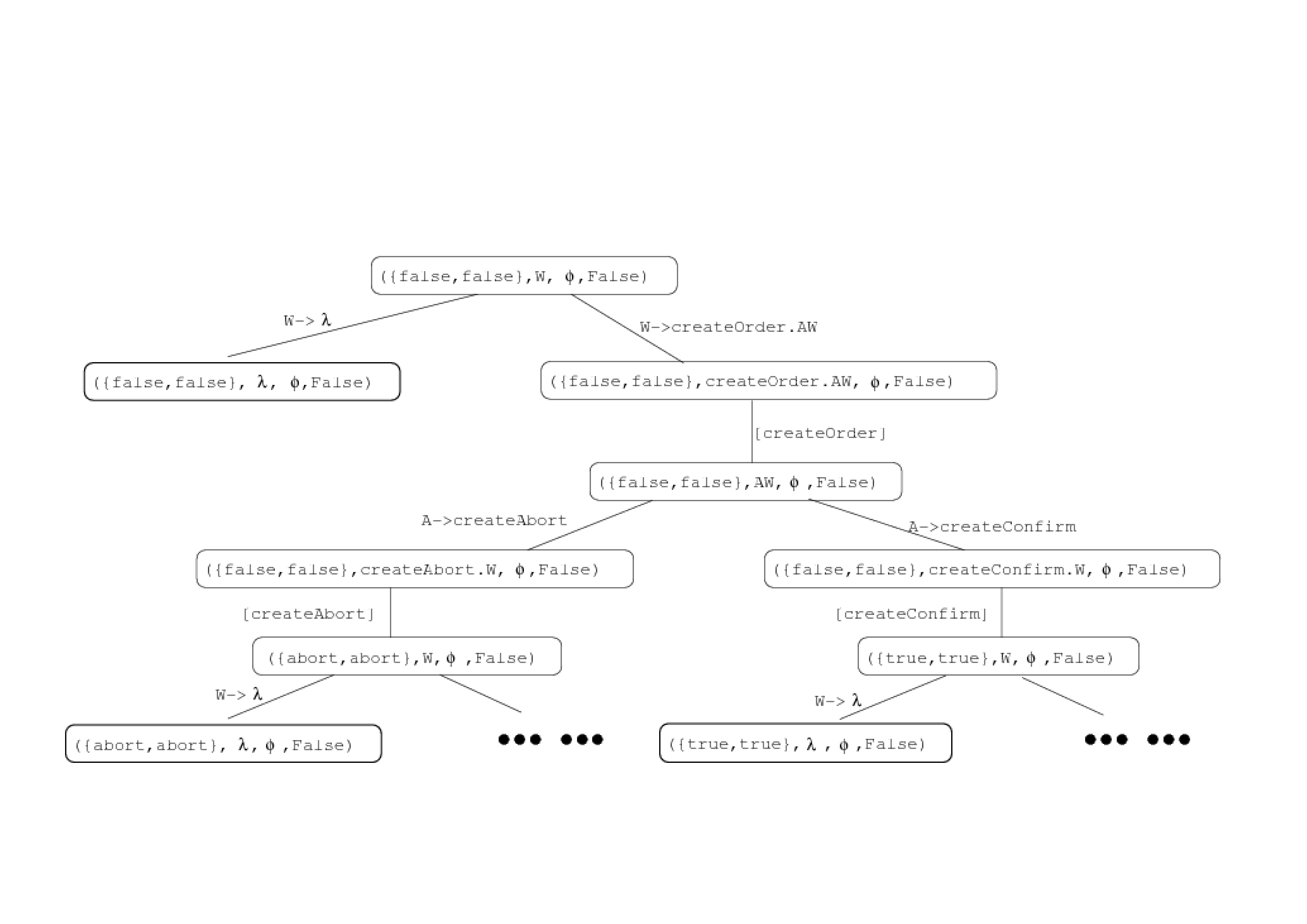}\\
\caption{A PLAY-tree for the web order system in Example~\ref{ex:webService1}}
\label{fig:ws1}
\end{figure*}

\noindent
Figure~\ref{fig:ws1} shows a PLAY-tree for the web service with the above CFG input,
where the round box and the bold one denote an internal node and an leaf node, respectively.
The object state of the web service system is simply represented as a set of properties
{\tt \{RBC.conf, STC.conf\}}. Thus, the root node is the ID
{\tt (\{false,false\},W,$\emptyset$,False)}, where both parties initially
set their local variable {\tt false}.
Even though the input regular expression contains infinite sequences,
for each finite sequence $w \in L(G)$, there is a {\em corresponding} success branch in the
PLAY-tree, along which if we concatenate all the labels of terminal moves
from the root, the result will be the finite message sequence $w$.
The PLAY-tree is not complete due to the
infinite branches.

Note that in each node of the PLAY-tree, the set
of current running LSCs is always $\emptyset$. That is because in this web order system example,
(i)~no LSCs are active initially; (ii) after receiving an external event
and then processing all enabled internal events,
all the activated LSCs will be complete before receiving a next external event. This is consistent to the simulation results in the super-step mode
using the Play-engine~\cite{HM03}. Also, we choose this example particularly
to leave the details in~\cite{HM03} on how exactly the transition function $\Delta$ works, due to the space limitation.

It needs to be mentioned that for a finite sequence $w \in L(G)$,
there may exist multiple corresponding branches, each of which may be a
success or violating branch. There could be two different reasons.
One possibility is that the grammar $G$ for an input language may be
ambiguous, thus causing two different derivations for the same sequence.
However, such an ambiguity can be avoided since we will constraint our input
language on regular language, which can be represented by an unambiguous grammar. The other is due to the fact that for a terminal move
\[
 (Q_1, aW, \mathcal{RL}_1, B_1) \vdash (Q_2, W, \mathcal{RL}_2, B_2),
\]
there could be multiple different $(Q_2, \mathcal{RL}_2, B_2) \in \nabla((Q_1, \mathcal{RL}_1, B_1), a)$
due to the nondeterministic nature of the LSC specification (e.g., two
internal events could be executed in different orders, but
result in different next system states).



\begin{figure}

  \begin{center}
    \includegraphics[height=1.6in,width=2.4in]{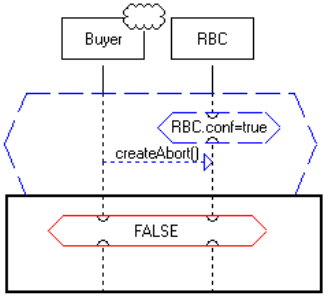}
  \end{center}

  \caption{An Anti-Scenario LSC}
  \label{fig:anti}
\end{figure}

\begin{Ex}
  Consider a scenario in the web order example, where the buyer wants
  to abort the order after the order has been confirmed. Let's assume
  this scenario as a forbidden one, and model it in LSC as an anti-scenario LSC
  as shown in Figure~\ref{fig:anti}.
  Consider the web order system with LSCs in Figure~\ref{fig:webService} and~\ref{fig:anti},
  as well as the input $(createOrder+createAbort+createConfirm)^{*}$,
  which can be transformed to a context-free grammar
  $G_1$ as follows:
  \[
    W \rightarrow \lambda\ \mid\ createOrder\cdot W \ \mid\ createAbort\cdot W \ \mid createConfirm \cdot W
  \]
\label{ex:webService2}
\end{Ex}

It is easy to find out that there exists a failure branch
in its corresponding PLAY-tree\footnote{We omit its PLAY-tree
here due to the space limitation.} leading to the anti-scenario.
We call the event sequence in a failure branch
a {\em prefix} of failure trace. A prefix trace is a least one if
there is no strict prefix of this trace leading to a failure leaf
in the PLAY-tree. Hence, for this example,
$createOrder\cdot createConfirm\cdot createAbort$ is
a least prefix of a failure trace causing the anti-scenario.

\subsection{A Memoized Depth-first Search}

Let an LSC model be a three-tuple $(Sys, Ls, G)$, where
$Sys$ is a reactive system, $Ls$ is its LSC specification,
and $G$ is a context-free grammar denoting the input language.
The consistency of an LSC model $(Sys, Ls, G)$ can be defined in terms
of a PLAY-tree. It is consistent if all
the finite branches in the PLAY-tree are success branches;
otherwise, the LSC model
is inconsistent if there exists a failure branch in the PLAY-tree;
and a prefix of failure trace can be
easily located along the failure branch. Therefore,
the web order system with the CFG $G$
in Example~\ref{ex:webService1} is consistent because
for each finite sequence $w \in L(G)$, there is a
{\em corresponding} success branch in the PLAY-tree
shown in Fig.~\ref{fig:ws1};
while the same system with the anti-scenario and the CFG $G_1$
is inconsistent due to the existence of a failure branch.

Given an LSC model $(Sys, Ls, G)$, the consistency checking is basically
a traversal of its PLAY-tree to see whether there exists any failure branch.
Due to the recursion nature of the context-free grammar $G$,
there may exist many infinite branches in the PLAY-tree, or
the finite branch could be any long.
Therefore, neither depth-first nor breadth-first strategy
is good enough for the completion of consistency check over
the PLAY-tree.

We introduce a memoized depth-first search strategy for
traversing a PLAY-tree, such that
any repeated IDs seen later during the traversal will not
be explored again. A memoized depth-first strategy is
essentially an extension of standard depth-first search
where visited IDs are recorded so that their later occurrences
can be simply considered a cycle of moves.

\begin{figure}[htp]
\begin{quote}
{
\small\tt

1.~~$GT$: global variable to record seen IDs\\
2.~~{\bf (Bool, $\Sigma^{*}$)} $mdft$($(Q, W, R, B)$:ID, $Tr$:$\Sigma^{*}$)\newline
3.~~{\bf if} ($B$ is {\bf True}) \\
4.~~\verb+  +{\bf return} ({\bf True},$Tr$);~~\%\% violation\\
5.~~{\bf if} ($W$ is $\lambda$)\\
6.~~\verb+  +{\bf return} ({\bf False},$Tr$);~~\%\% success\\
7.~~{\bf if} ($(Q, W, R, B) \in GT$)\\
8.~~\verb+  +{\bf return} ({\bf False},$Tr$);~~\%\% memoized ID\\
9.~~$GT \leftarrow GT \cup \{(Q,W,R,B)\}$;\\
10.~{\bf Let} $W = AW_1$;~~\%\% $A$ is the first char\\
11.~{\bf if} ($A$ is an external event) \\
12.~\verb+  +$Tr \leftarrow A + Tr$; \%\% concatenation\\
13.~\verb+  +{\bf for} each $(Q_1,R_1,B_1) \in \nabla((Q,R,B),A)$\\
14.~\verb+    +$(V,Tr1) \leftarrow mdft((Q_1,W_1,R_1,B_1),Tr)$;\\
15.~\verb+    +{\bf if} ($V$ is {\bf True})\\
16.~\verb+      +{\bf return} $(V,Tr1)$;\\
17.~{\bf else}~~\%\% if $A$ is a nonterminal variable\\
18.~\verb+  +{\bf for} each $A \rightarrow D_1\cdots D_n$, where $n \geq 0$\\
19.~\verb+    +$(V,Tr1) \leftarrow mdft((Q,D_1\cdots D_nW_1,R,B),Tr)$;\\
20.~\verb+    +{\bf if} ($V$ is {\bf True})\\
21.~\verb+      +{\bf return} $(V,~Tr1)$;\\
22.~{\bf return} ({\tt False}, $Tr$);
}
\end{quote}
\caption{An algorithm for a memoized depth-first traversal of PLAY-tree}
\label{fig:mdft}
\end{figure}

Figure~\ref{fig:mdft} shows a pseudo code algorithm for consistency checking,
where the recursive function {\em mdft}
takes an initial ID and an initially empty trace as input,
and checks in a memoized depth-first order whether the LSC
specification is consistent over all
the input event sequences.
The function {\em mdft} returns a pair {\tt\em (B,Trace)}
consisting of a violation indicator {\tt\em B} and a failure
trace {\tt\em Tr} if violated.
A global table, named {\em GT}, is
introduced to record all the IDs in the visited internal nodes
(line 9). If a new node has an ID which has already
been in the global table, then no exploration will be taken below
this node (line 7-8); otherwise, appropriate moves will take place
depending on the leftmost symbol in the sequence input $W$ (lines 10-22).
The local variable $Trace$ is used to concatenate all the external
events along the current branch from root to
the current calling ID (line 12).
Whenever a violation ID is found, it will immediately return the failure
$Trace$ to the parent caller (lines 3-4, 15-16, 20-21);
otherwise, the function $mdft$ will check all input event sequences
to make it sure that the LSC system is consistent.

Different from the automaton-based strategies~\cite{HK99,BH01,KTWW06,KM08}
that the LSC system is transformed into an equivalent automaton, in which the
language of the LSC model is defined, our work here designs a
computational scheme and implements a logic-based simulator to show whether
an LSC system is consistent to a given language defined over a set of external events.
Additionally, our L2C2 system provides a great flexibility to extend the LSC
simulator with new language features and functionalities.

\section{External Events Specification Language (EESL)}
\label{sec:eesl}

We define a language, called {\em EESL}, for users to specify a set of external
event sequences for checking whether a reactive system is consistent given each
sequence as an input. The EESL, as shown below, basically extends the regular
expression notation with a parallel operator $\|$ and
a testing operator $\langle\rangle$.
\begin{eqnarray*}
S & ::= & S+S\ |\ S\cdot S\ |\ {S}^{*}\ |\ (S)\ |\ S \| S\ |\ a\ |\ \langle a \rangle  \\
 & & \mbox{for any }a \in \Sigma, \mbox{ the set of external events}
\end{eqnarray*}

\subsection{The Parallel Operator $\|$}

Both our L2S simulator and the PLAY-engine~\cite{HM03}
assume an input-enabled concurrency model, where no two
external events occur at the same time, and that before processing
a next external event, the simulated system always reaches a stable state,
where all enabled internal events have been processed.
However, this raises a problem especially in a distributed web service system,
where multiple external events may come from different service points in parallel.
Consider the web order example shown in Figure~\ref{fig:webService} again.
Once an order is initiated by a Buyer, for a period of time,
the Buyer can create an abort request, or the seller can confirm
the order, or both happen in parallel.
However, under the current system setting, it is not easy
to simulate the above scenario with parallel external events.

\begin{figure*}[!ht]
	\centering
	\begin{tabular}{cc}
		\includegraphics[width=2.4in,height=2in]{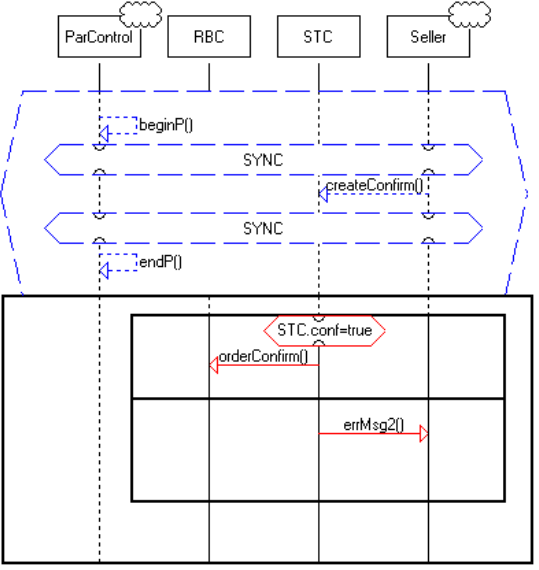}
		&
		\includegraphics[width=2.4in,height=2in]{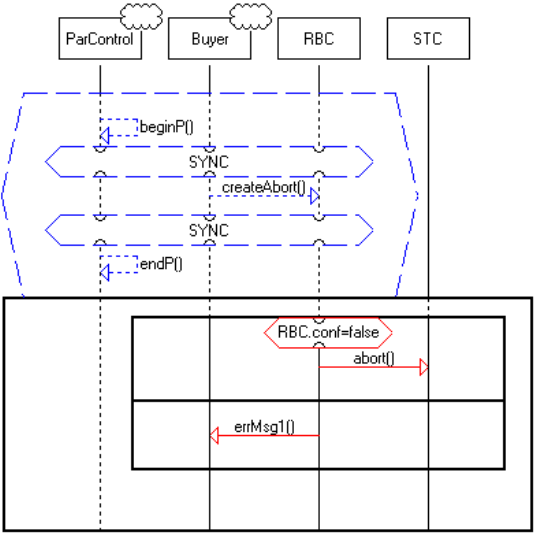} \\
	  (a) Seller creates an order confirmation
	  &
	  (b) Buyer creates an abort request \\
	
    \multicolumn{2}{c}{
	  \includegraphics[width=2in,height=1.2in]{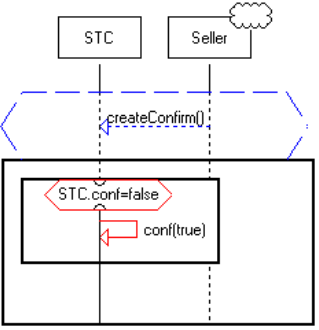}} \\
		 \multicolumn{2}{c}{
		  (c) STC sets the confirmation status}
	\end{tabular}
\caption{An LSC specification with a Parallel object}
\label{fig:ws2}
\end{figure*}

We introduce a parallel operation $a \| b$ to support
interleaved parallel external events.
At the same time, we extend the LSC with a virtual parallel object,
{\tt ParControl}, with two functional controls {\em beginP} and {\em endP},
which denote the beginning
and the end, respectively, of a scenario for receiving
interleaved parallel external events. A parallel operation
$a\|b$ can thus be transformed to $beginP\cdot(a+b+ab+ba)\cdot endP$ in regular
expression. Figure~\ref{fig:ws2} illustrate the use of {\tt ParControl} in
the web order example. {\tt SYNC} is only used to synchronize the
instance locations, making it sure that the external events {\em createConfirm} and
{\em createAbort} may happen between the parallel controls {\em beginP} and
{\em endP}, so that the events in the main charts of LSCs~\ref{fig:ws2}(a)
and~\ref{fig:ws2}(b) can be simulated concurrently after the common
virtual message {\em endP}. Both $beginP$ and $endP$ are implicitly
generated by the simulator to enable the interleaved parallelism among
external events.

Note that the LSC (g) in Figure~\ref{fig:webService} has been
broken into two new LSCs (a) and (c) in Figure~\ref{fig:ws2},
so that LSC~\ref{fig:ws2}(c) will be performed before {\em endP},
unparallel with the main chart of LSC~\ref{fig:ws2}(b).
The events defined in the main chart of LSC~\ref{fig:ws2}(c)
actually forms an atomic block, which will be executed uninterruptedly.
Otherwise, without this atomic block, the web order system possibly leads to
a situation where RBC aborts the order while STC confirms the order.

As a scenario-based programming language, LSC has a distinguished nature
to model the concurrency for distributed reactive systems;
it would be much better and easier (e.g., to model our web order example)
if the LSC language also provided a function to define an atomic block.
Atomicity in concurrent systems using LSCs will be explored in our future work.

In the rest of paper, whenever the web order example is mentioned,
we use the LSCs in Figure~\ref{fig:ws2} to substitute the original
LSCs (c) and (g) in Figure~\ref{fig:webService}.

\subsection{The Testing Operator $\langle\rangle$}

\begin{figure*}[!ht]
	\centering
	\begin{tabular}{cl}
		\includegraphics[width=2.4in,height=2in]{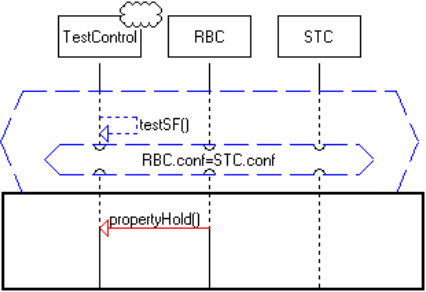}
		&
		\includegraphics[width=2.4in,height=2in]{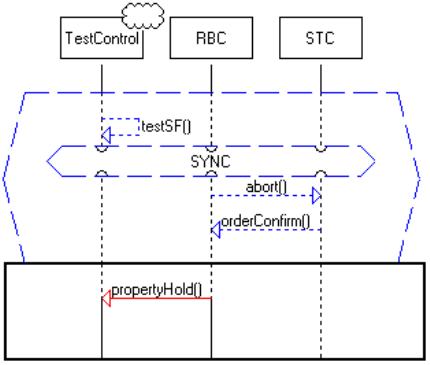} \\
	  (a) RBC.conf is same as STC.conf
	  &
	  (b) STC receives an abort message but still\\
	  & ~~~~ send an order confirmation.
	\end{tabular}
\caption{State and Path predicates specification}
\label{fig:ws3}
\end{figure*}

We introduce a testing operator $\langle\rangle$ as well as a virtual
testing object, {\tt testControl}, in LSCs  for users to specify
and test certain temporal properties. $\langle a\rangle$ is automatically
transformed to a regular expression $testSF\cdot a$, where $a$ is
an external event, and $testSF$ is an event to trigger the
virtual testing object, {\tt testControl}, in LSCs and
thus activate testing LSC scenarios.
Both a state predicate, denoting whether a predicate is true
in a certain ID point, and a path (or trajectory) predicate,
denoting whether a predicate will be true from a certain ID point,
can be specified using a testing LSC scenario.
For example, the LSC in Figure~\ref{fig:ws3}(a) defines a
state predicate checking whether the condition of
``$RBC.conf=STC.conf$'' is true, and the one in Figure~\ref{fig:ws3}(b)
defines a path predicate checking whether the scenario,
that the STC receives an abort message but then still
send an order confirmation, will happen in the future.

The testing operator $\langle\rangle$ and the object {\em testControl},
incorporated with our computational PLAY-tree, provides
an easy vehicle to modeling and testing CTL properties in running LSCs.
Our L2C2 system provides a special testing mode to run the LSC model,
as well as checking the satisfiability of the specified testing
predicates through all paths of the computational PLAY-tree.
In the testing mode, the L2C2 system will automatically
produce the testing trigger {\em testSF} right before generating
each single external event or each occurrence of interleaved parallel events,
so that the state and the path predicates in testing LSCs will be activated
and checked at each super state, that is, the initial state or
a stable state after completely processing an external event or a parallel operation.
The special event {\em propertyHold} to the object {\tt testControl},
as shown in Figure~\ref{fig:ws3}, can be detected by the L2S simulator, indicating
that the testing scenario has been satisfied. All the testing results
with CTL operators will be reflected later in the justification section.

\subsection{From Regular Expression to Context-Free Production Grammar}

The testing language on external events is
given by users in EESL, which is essentially a notation of
regular expression since both $\|$ and $\langle\rangle$
can be represented in regular expression.
The L2C2 system automatically transform the language input in EESL
into an executable right recursive context-free production grammar (CFPG)
in definite logic clauses.
The CFPG is used to generate all possible testing event sequences for
the LSC simulator.

\section{Justification}
\label{sec:justification}

Given an LSC specification and an input language in EESL,
the LSC simulator returns a truth value as well as its justification, which provides evidences
so that users can easily follow the
evidences to re-establish the simulation scenarios.

\begin{figure*}[!ht]
	\centering
	\begin{tabular}{cl}
		\includegraphics[width=2in,height=2in]{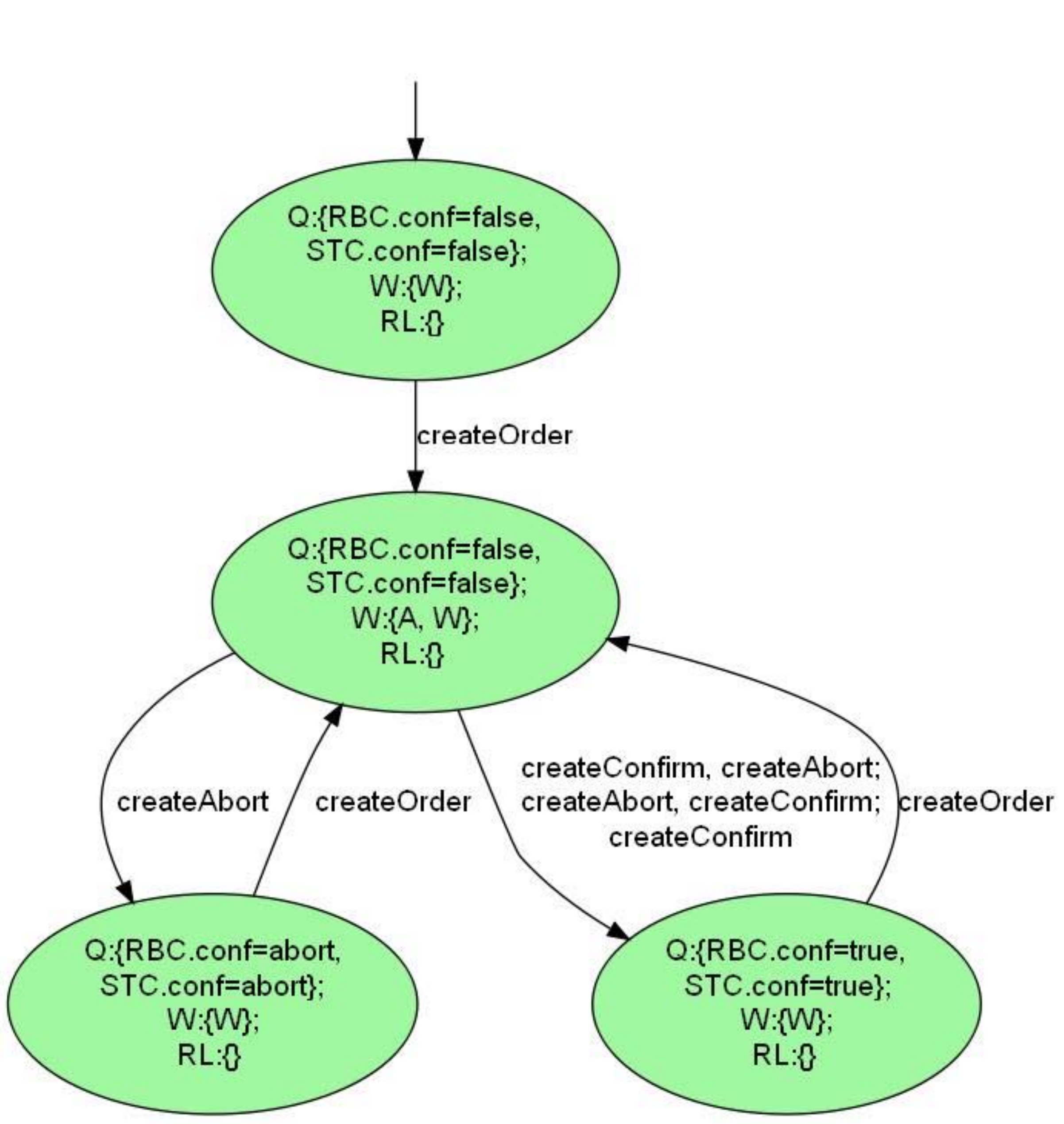}
		&
		\includegraphics[width=2.8in,height=2in]{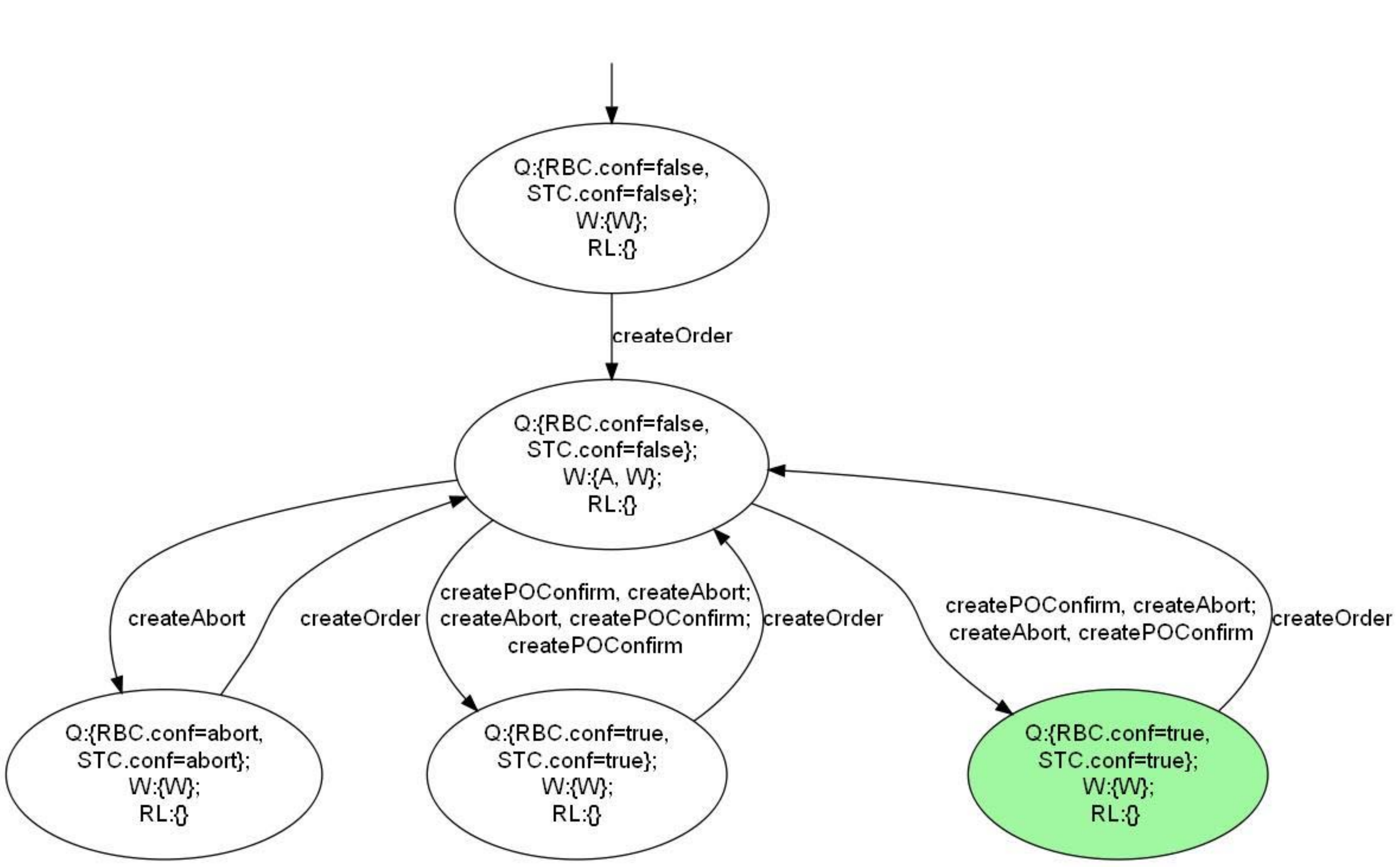} \\
	  (a) Saftyness: AG$(RBC.conf=STC.conf)$
	  &
	  (b) Reachability: EF$(abort\wedge orderConfirm)$
	\end{tabular}
\caption{ID predicate specification}
\label{fig:ws4}
\end{figure*}

If the consistency is false, it returns a least prefix failure trace
as an evidence; while if the result is true, then
it actually returns a state transition graph, showing labeled transitions
among states after processing each external event and the testing temporal properties.
Figure~\ref{fig:ws4} shows two positive justification graphs,
where the green (or gray) color represents the satisfiability of a testing
scenario. Each node in the graph represents a super state, either an initial state or
a stable state after processing an external event or parallel external events;
the text in each node shows the current ID; the labels on a transition
mean the external events, separated by a semi-colon; a label of two external
events (e.g., ``a,b''), separated by a comma, represents interleaved parallel
events in this order.

Figure~\ref{fig:ws4}(a) verifies the saftyness property
in CTL: {\tt AG}$(RBC.conf=STC.conf)$, where the state
predicate is globally true on every path.
Figure~\ref{fig:ws4}(b) verifies the reachability property
in CTL: {\tt EF}$(abort\wedge orderConfirm)$,
which means that it is possible that in some scenario,
the STC receives an abort event but sends an orderConfirm event to
RBC. Users can also re-construct the simulation scenarios using the
justification evidence in an interactive way with PLAY-engine.

\section{Conclusion}
\label{sec:conclusion}

We introduce a logic-based LSC consistency checking system,
named {\em L2C2}. An LSC simulator, named {\em L2S},
has been presented, utilizing a memoized depth-first search strategy
over a computational PLAY-tree, to show how a reactive system in LSCs
would response to a set of external event sequences.
A formal language, named {\em EESL}, has been defined
to specify external event sequences. The language extends
a regular expression notation with a parallel operator $\|$
and a testing operator $\langle\rangle$.
The parallel operator, combined with a virtual parallel control in LSCs,
allows interleaved parallel external events to be processed in LSCs.
The testing operator, combined with a virtual testing control in LSCs,
provides users to a new approach to specify and test CTL temporal properties.
The L2C2 system further provides either a state transition graph or
a failure trace to justify the consistency checking results.
We believe that an automatic LSC simulator as well as
debugging, temporal property testing, and consistency checking
capabilities will play important roles on designing trustworthy
reactive distributed software and hardware processes.

\bibliographystyle{latex8}

\begin{thebibliography}{10}

\bibitem{Abramson84}
Harvey Abramson:
\newblock Definite Clause Translation Grammars.
\newblock {\em International Symposium on Logic Programming}, pp. 233-240., 1984.

\bibitem{BGS05}
A. Bunker, G. Gopalakrishnan, and K. Slind:
\newblock Live Sequence Charts Applied to Hardware Requirements Specification and Verification:
A VCI Bus Interface Model.
\newblock {\em Software Tools for Technology Transfer}, 7(4):341--350, 2005.


\bibitem{BH01}
Y. Bontemps, P. Heymans:
\newblock Turning high-level live sequence charts into automata.
\newblock {\em Proceedings of Scenarios and State-Machines: Models, Algorithms,
and Tools}, 2002.

\bibitem{BHK03}
Y. Bontemps, P. Heymans, H. Kugler:
\newblock Applying LSCs to the specification of an air traffic control system.
\newblock {\em Workshop on Scenarios and State Machines: Models, Algorithms and Tools},
2003.

\bibitem{BDK02}
J. Bohn, W. Damm, J. Klose, A. Moik, and H. Wittke:
\newblock Modeling and validating train system applications using statemate
and live sequence charts.
\newblock {\em The 6th Biennial World Conference on Integrated Design and Process Technology},
2002.


\bibitem{CGP01}
E.M. Clarke, O. Grumberg, and D.A. Peled:
\newblock {\em Model Checking}.
\newblock The MIT Press, 2001.

\bibitem{CHK05}
P. Combes, D. Harel, and H. Kugler:
\newblock Modeling and Verification of a Telecommunication Application using
Live Sequence Charts and the Play-Engine Tool.
\newblock {\em The 3rd Int. Symp. on Automated Technology for Verification and Analysis},
2008.


\bibitem{DH99}
Werner Damm and David Harel:
\newblock LSCs: Breathing Life into Message Sequence Charts.
\newblock {\em Proc. 3rd IFIP Int. Conf. on Formal Methods for Open Object-based Distributed Systems},
pp. 293--312, 1999.


\bibitem{H00}
David Harel:
\newblock From Play-In Scenarios to Code: An Achievable Dream.
\newblock {\em Proc. Fundamental Approaches to Software Engineering (FASE)},
pp. 22--34, 2000.

\bibitem{HK99}
David Harel and Hillel Kugler:
\newblock Synthesizing State-based object systems from LSC specifications.
\newblock {\em Int. Journal of Foundations of Computer Science},
13(1): 5--51, 2002.

\bibitem{HM03}
David Harel and Rami Marelly:
\newblock {\em Come, Let's Play: Scenario-Based Programming Using LSCs and the Play-Engine}.
\newblock Springer-Verlag, 2003.

\bibitem{HMS08}
D. Harel, S. Maoz, and I Segall:
\newblock Some results on the expressive power and complexity of LSCs.
\newblock LNCS 4800, pp. 351-366, 2008.


\bibitem{KH05}
H. Kugler, D. Harel, A. Pnueli, Y. Lu, and Y. Bontemps:
\newblock Temporal logic for scenario-based specifications.
\newblock In TACAS 2005, pp. 445--460.


\bibitem{KM07}
R. Kumar and E. Mercer:
\newblock Improving translation of live sequence charts to temporal logic.
\newblock {\em Int. conf. on automated verification of critical systems},
pp. 183--197, 2007.

\bibitem{KM08}
R. Kumar and E. Mercer:
\newblock Improving live sequence chart to automata translation for verification.
\newblock {\em Electronic Communications of the EASST}, 2008.



\bibitem{KTWW06}
J. Klose, T. Toben, B. Westphal, and H. Wittke:
\newblock Check It Out: On the Efficient Formal Verification of Live Sequence Charts.
\newblock {\em 18th International Conference on Computer Aided Verification} (CAV), pp. 219--233, 2006.

\bibitem{SD05}
Jun Sun and Jin Song Dong:
\newblock Model checking live sequence charts.
\newblock {\em The 10th IEEE int. conf. on Engineering of complex computer systems}, 2005.


\bibitem{TW06}
T. Toben and B. Westphal:
\newblock On the expressive power of LSCs.
\newblock {\em The 32nd Conf. on Current Trends in Theory and Practice of Computer Science}, pp. 33--43, 2006.

\bibitem{UMLdocs}
The Object Management Group (OMG):
\newblock Documentation of the Unified Modeling Language.
\newblock http://www.omg.org.

\bibitem{UML2}
UML:
\newblock {\em Unified Modeling Languages Superstructure Specification}, v2.0.
\newblock http://www.uml.org/, OMG specification, 2005.

\bibitem{Z120}
Z.120 ITU-T Recommendation:
\newblock {\em Message Sequence Chart (MSC)}.
\newblock ITU-T, 1996.


\end{thebibliography}


\end{document}